\documentclass[3p,times,twocolumn]{elsarticle}

\usepackage{ecrc}

\volume{00}

\firstpage{1}

\journalname{Nuclear Physics B Proceedings Supplement}

\runauth{P.S.B.~Dev, P.~Millington, A.~Pilaftsis, D.~Teresi}

\jid{nuphbp}

\jnltitlelogo{Nuclear Physics B Proceedings Supplement}

\usepackage{amssymb}
\usepackage[figuresright]{rotating}

\usepackage{bm, dsfont,
  amssymb, amsmath,
  graphicx,
  hyperref,
  color,
  braket,
  array,
  slashed,
  mathrsfs,
  subfig,
  upgreek}
\usepackage[font={footnotesize}]{caption}

\DeclareMathAlphabet{\mathpzc}{OT1}{pzc}{m}{it}
\DeclareMathOperator{\Tr}{Tr}
\newcommand{\mat}[1]{\bm{#1}}

\newcommand{\ve}[1]{\mathbf{#1}}

\newcommand{\D}[2]{\mathrm{d}^{#1}{#2}}

\newcommand{\lsim}{\lesssim}

\newcommand{\CP}{C\!P}
\newcommand{\gCP}{\widetilde{C}\!P}

\newcommand{\Edu}[4]{[E_{#1}(\ve #2)]_{#3}^{\phantom{#3} #4}}
\newcommand{\Esud}[4]
{\left[(2 E_{#1}(\ve #2))^{-{1}/{2}}\right]^{#3}_{\phantom{#3}#4}}
\newcommand{\Esdu}[4]
{\left[\big(2 E_{#1}(\ve p)\big)^{-\frac{1}{2}}\right]_{#3}^{\phantom{#3}#4}}

\newcommand{\su}[4]{[u(\ve #2, #1)]_{#3}^{\phantom #3 #4}}
\newcommand{\sub}[4]{[\bar{u}(\ve #2, #1)]^{#3}_{\phantom #3 #4}}
\newcommand{\sv}[4]{[v(\ve #2,#1)]_{#3}^{\phantom #3 #4}}

\newcommand{\n}[6]{[n_{#2}^{#1}(\ve #3,#6)]_{#4}^{\phantom{#4}#5}}
\newcommand{\nb}[6]
{[\bar{n}_{#2}^{#1}(\ve #3,#6)]_{#4}^{\phantom{#4}#5}}

\newcommand{\DiT}{(2 \pi)^3 \delta^{(3)}}
\newcommand{\DiFud}[5]
{{(2 \pi)^4 [\delta^{(4)}(#1)]^{#2 \phantom{#3 #4} #5}
    _{\phantom #2 #3 #4}}}

\newcommand{\edu}[4]{\big[e^{#1 i #2 \cdot x}\big]_{#3}
  ^{\phantom{#3} #4}}

\newcommand{\h}[2]{h_{#1}^{\phantom{#1}#2}}
\newcommand{\hs}[2]{h_{\phantom{#1}#2}^{#1}}
\newcommand{\hr}[2]{\mathbf{h}_{#1}^{\phantom{#1}#2}}

\newcommand{\hrc}[2]{[\mathbf{h}^{\tilde{c}}]^{#1}
  _{\phantom{#1}#2}}

\newcommand{\Tdu}[5]
{{#1}_{#2 \phantom{#3} #4 \phantom{#5}}
  ^{\phantom{#2} #3 \phantom{#4}  #5}}

\usepackage{tikz}
\newcommand{\boxalign}[2][0.97\textwidth]{
 \par\noindent\tikzstyle{mybox} = [draw=black,inner sep=6pt]
 \begin{center}\begin{tikzpicture}
  \node [mybox] (box){%
   \begin{minipage}{#1}{\vspace{-5mm}#2}\end{minipage}
  };
 \end{tikzpicture}\end{center}
}

\usepackage{textpos}
\begin{document}

\begin{frontmatter}

%% Title, authors and addresses

%% use the tnoteref command within \title for footnotes;
%% use the tnotetext command for the associated footnote;
%% use the fnref command within \author or \address for footnotes;
%% use the fntext command for the associated footnote;
%% use the corref command within \author for corresponding author footnotes;
%% use the cortext command for the associated footnote;
%% use the ead command for the email address,
%% and the form \ead[url] for the home page:
%%
%% \title{Title\tnoteref{label1}}
%% \tnotetext[label1]{}
%% \author{Name\corref{cor1}\fnref{label2}}
%% \ead{email address}
%% \ead[url]{home page}
%% \fntext[label2]{}
%% \cortext[cor1]{}
%% \address{Address\fnref{label3}}
%% \fntext[label3]{}

\dochead{\hfill \small MAN/HEP/2014/12; IPPP/14/45; DCPT/14/90}
%% Use \dochead if there is an article header, e.g. \dochead{Short communication}

\title{Flavour Covariant Formalism for Resonant Leptogenesis}

\author[a]{P.~S.~Bhupal Dev \fnref{1}}
%\ead{Bhupal.Dev@hep.manchester.ac.uk}

\author[a,b]{Peter Millington}
%\ead{Peter.Millington@manchester.ac.uk}

\author[a]{Apostolos Pilaftsis}
%\ead{Apostolos.Pilaftsis@manchester.ac.uk}

\author[a]{Daniele Teresi}
%\ead{Daniele.Teresi@hep.manchester.ac.uk}

\address[a]{~Consortium for Fundamental Physics,
  School of Physics and Astronomy, \\ 
  University of Manchester, Manchester M13 9PL, United Kingdom.\smallskip}

\address[b]{ ~Institute for Particle Physics Phenomenology, 
  Durham University, Durham DH1 3LE, United Kingdom.}

\fntext[1]{Speaker.}

\begin{abstract}
We present a fully flavour-covariant formalism for transport phenomena and apply it to study the flavour-dynamics of  Resonant Leptogenesis (RL). We show that this formalism provides a complete and unified description of RL, consistently accounting for three {\em distinct} physical phenomena: (i) resonant mixing and (ii) coherent oscillations between different heavy-neutrino flavours, as well as (iii)  quantum decoherence effects in the charged-lepton sector. We describe the necessary emergence of higher-rank tensors in flavour space, arising from the unitarity cuts of partial self-energies. Finally, within a Resonant $\tau$-Genesis model, we illustrate the importance of this formalism by showing that the final lepton asymmetry can vary by as much as an order of magnitude between partially flavour-dependent treatments, which do not capture all of the pertinent flavour effects.
\end{abstract}


\end{frontmatter}

%\begin{textblock}{4}(12,-14.5)
%\begin{flushright}
%MAN/HEP/2014/12, IPPP/14/45, DCPT/14/90
%\end{flushright}
%\end{textblock}

%%
%% Start line numbering here if you want
%%
% \linenumbers

%% main text

%%%%%%%%%%%%%%%%%%%%%%%%%%%%%  %%%%%%%%%%%%%%%%%%%%%%%%%%%%%%%%%%%%%%%

\section{Introduction} \label{sec:1}
Leptogenesis~\cite{Fukugita:1986hr} is an elegant framework for dynamically generating the observed matter-antimatter asymmetry in our Universe through out-of-equilibrium decays of heavy Majorana neutrinos, whilst simultaneously explaining the smallness of the light neutrino masses by the 
seesaw  mechanism~\cite{seesaw}. Resonant Leptogenesis (RL)~\cite{Pilaftsis:1997dr,  Pilaftsis:2003gt} offers the possibility of realizing this beautiful idea at energy scales accessible to laboratory experiments. In RL, the   heavy  Majorana neutrino   self-energy
effects  on  the  leptonic $  \CP$-asymmetry  become
dominant~\cite{Flanz:1994yx} and get resonantly enhanced, when 
at least two of the heavy  neutrinos have a  small mass difference  
comparable to their  decay  widths~\cite{Pilaftsis:1997dr}.  

Flavour effects in both heavy-neutrino and charged-lepton sectors, as well as the interplay between them, play an important role in determining the final
lepton asymmetry in low-scale leptogenesis models~\cite{Abada:2006fw, Nardi:2006fx}. These intrinsically quantum effects  can be consistently accounted for by extending
the classical  flavour-diagonal Boltzmann equations for the number  densities of individual
flavour species to a  semi-classical evolution equation for a
{\it matrix  of  number densities}~\cite{Sigl:1993}. Using this general technique,  we present  in Section~\ref{sec:2} a {\it fully}  flavour-covariant formalism for transport phenomena in the Markovian regime. As an application of this general formalism, we derive  a set of flavour-covariant transport
equations  for lepton and heavy-neutrino  number densities  with arbitrary flavour content in a quantum-statistical ensemble. 
We demonstrate  the necessary  appearance of  rank-4 tensor
rates  in  flavour  space  that  properly  account  for  the  statistical
evolution of  off-diagonal flavour coherences. 
As shown in Section~\ref{sec:3}, this manifestly flavour-covariant formalism enables
us to capture three important flavour effects pertinent to RL: (i) the
resonant  mixing of  heavy neutrinos,  (ii) the  coherent oscillations
between heavy neutrino flavours and (iii) quantum (de)coherence effects
in the charged-lepton sector. In Section~\ref{sec:4}, we present a numerical example to illustrate  the  importance of these flavour
off-diagonal effects on the final lepton asymmetry. Our conclusions are given in Section~\ref{sec:5}. 
For a detailed discussion of the topics presented here, we refer the reader to~\cite{Dev:2014laa}. 

%%%%%%%%%%%%%%%%%%%%%%%%%%%%% PM %%%%%%%%%%%%%%%%%%%%%%%%%%%%%%%%%%%%%%%

\section{Flavour-Covariant Formalism} \label{sec:2}

Let us begin with an arbitrary flavour content for the lepton doublet field operators $L_l$ (with $l=1,\: 2,\: \dots,\:  \mathcal{N}_{L}$) and the right-handed Majorana neutrino field operators $N_{\rm R, \alpha}  \equiv \mathrm{P_R}  N_\alpha$ (with $\alpha=1,\: 2,\: \dots,\: \mathcal{N}_N$), where $\mathrm{P_R} = (\mat{1}_4 + \gamma_5)/2$ is the 
right-chiral projection  operator. The field operators transform as follows in the fundamental representations of $U(\mathcal{N}_{L})$ and $U(\mathcal{N}_{N})$: 
\begin{subequations}
\begin{gather}
\hspace{-1.0em}L_l \to L'_l = V_l^{\phantom{l}m}L_m\;,\quad 
L^l \equiv (L_l)^{\dag} \to L'^l = V^l_{\phantom{l}m}L^m\;, 
\\
\hspace{-1.0em}
N_{\mathrm{R},\, \alpha} \to N'_{\rm R, \alpha} = U_{\alpha}^{\phantom{\alpha}\beta}N_{\mathrm{R},\,\beta},\quad 
N_{\mathrm{R}}^{\alpha} 
\to 
N_{\rm R}'^{ \alpha} =  U^{\alpha}_{\phantom{\alpha}\beta}N_{\mathrm{R}}^{\beta},
\end{gather}
\end{subequations}
where $V_l^{\phantom{l}m} \in U(\mathcal{N}_{L})$ and $U_{\alpha}^{\phantom{\alpha}\beta} \in U(\mathcal{N}_{N})$. In the flavour basis, the relevant neutrino Lagrangian is given by 
\begin{equation}
  -\mathcal{L}_N   =  \h{l}{\alpha}  \overline L^{l} 
  \widetilde{\Phi}  N_{\rm R, \alpha} 
  + \frac{1}{2} \overline{N}_{\rm R, \alpha}^C  [M_N]^{\alpha \beta} 
  N_{\rm R, \beta} + {\rm H.c.}\;,
  \label{eq:L}
\end{equation}
where $\widetilde{\Phi}=i\sigma_2\Phi^*$ is the isospin conjugate of the Higgs doublet $\Phi$. The Lagrangian~\eqref{eq:L} transforms covariantly under $U(\mathcal{N}_{L})\otimes U(\mathcal{N}_{N})$, provided the heavy-neutrino Yukawa and mass matrices transform as
\begin{subequations}
\begin{align}
  \h{l}{\alpha} \ &\rightarrow \ h_{l}'^{\ \alpha} \ = \ V_l^{\phantom l m}
  \;
  U^\alpha_{\phantom{\alpha} \beta} \; \h{m}{\beta} \; ,\\ 
  [M_N]^{\alpha \beta} \ &\rightarrow \ 
[M'_N]^{\alpha \beta} \ = \ U^\alpha_{\phantom{\alpha} \gamma} \;
  U^\beta_{\phantom{\beta} \delta} \; [M_N]^{\gamma \delta} \;.
\end{align}
\end{subequations}
The field operators in~\eqref{eq:L} can be expanded in flavour-covariant plane-wave decompositions, e.g.
\begin{align}
L_l(x) \ & = \  \sum_{s=+,-} \int_{\ve p} \Esdu{L}{p}{l}{i}
\nonumber \\ & \quad \times \ 
\Big( \edu{-}{p}{i}{j} \su{s}{p}{j}{k} \,
    b_k(\ve p,s,0)  \;
    \nonumber\\&\qquad +\: \edu{}{p}{i}{j} \, \sv{s}{p}{j}{k} \,
    d_k^{\dagger}(\ve p,s,0)
  \Big)\;,
\label{eq:Lfield}
\end{align}
where we have suppressed the isospin indices. In \eqref{eq:Lfield}, $\int_{\mathbf{p}}\equiv\int\!\frac{\D{3}{\mathbf{p}}}{(2\pi)^3}$, $s$ is the helicity index and $[E_L^2(\mathbf{p})]_{l}^{\phantom{l}m}=\mathbf{p}^2\delta_{l}^{\phantom{l}m}+
[M_L^{\dag}M_L]_{l}^{\phantom{l}m}$.
Notice that the Dirac four-spinors $\su{s}{p}{j}{k}$ and $\sv{s}{p}{j}{k}$ transform as rank-$2$ tensors in flavour space. The lepton creation and annihilation operators $b^k\equiv b_k^{\dag}$ and $b_k$, and the anti-lepton creation and annihilation operators $d^{\dagger,\, k}\equiv d_k$ and $d_k^{\dagger}$, satisfy the following equal-time anti-commutation relations
\begin{align}
  \label{eq:b_d_anticomm}
 & \big\{ b_l(\ve p,s,\tilde{t}), \,
  b^{m}(\ve p',s',\tilde{t}) \big\}  \  = \ \big\{d^{\dagger, m}(\ve p,s,\tilde{t}) , \,
  d_l^{\dagger}(\ve p',s',\tilde{t}) \big\} 
  \nonumber\\
  & \qquad \qquad \qquad  \qquad \  = \ \DiT{(\ve p - \ve p')} \, \delta_{s s'}\,
  \delta_l^{\phantom l m}  .
\end{align}
Note that for the Dirac field, the lepton annihilation operator $b_k(\ve p,s,\tilde{t})$ and the anti-lepton creation operator $d_k^{\dagger}(\ve p,s,\tilde{t})$ transform under the {\em same} representation of $U(\mathcal{N}_L)$. 

For the heavy {\em Majorana} neutrino creation and annihilation operators $a^{\alpha}(\ve k,r,\tilde{t})$ and $a_{\alpha}(\ve k,r,\tilde{t})$, with helicities $r=\pm$, it is necessary to introduce the flavour-covariant Majorana constraint
\begin{equation}
  d^{\dagger , \alpha}(\ve k,-\,r,\tilde{t})\ 
  = \ G^{\alpha \beta} \, b_\beta(\ve k,r,\tilde{t}) \ \equiv\ G^{\alpha\beta}a_{\beta}(\ve k,r,\tilde{t}) \;,
  \label{eq:def_G}
\end{equation}
where $G^{\alpha \beta}\equiv[ U^* U^\dagger ]^{\alpha \beta}$ are the elements of a unitary matrix $\bm{G}$, which transforms as a contravariant rank-$2$ tensor under $U(\mathcal{N}_N)$. Similar flavour rotations are {\it forced} by the flavour-covariance of the formalism, when we derive the transformation properties of the discrete symmetries $C$, $P$ and $T$. This necessarily leads to the {\it generalized} discrete transformations
\begin{subequations}
\begin{align}
 b_l(\ve p,s,\tilde{t})^{\widetilde{C}} \ & \equiv \ \mathcal{G}^{lm} \,
  b_m(\ve p,s,\tilde{t})^{C} \
  = \ -i \, d^{\dagger,l}(\ve p,s,\tilde{t}) \; ,\\
b_l(\ve p,s,\tilde{t})^P \ & 
  = \ - s \, b_l(-\ve p,-s,\tilde{t})\;, \\
 b_l(\ve p,s,\tilde{t})^{\widetilde{T}} \ &\equiv \
  \mathcal{G}_{lm} \, b_m(\ve p,s,\tilde{t})^{T} \
  = \   b_l(-\ve p,s,-\tilde{t}) \;, 
\end{align}
\end{subequations}
where $\mathcal{G}^{lm}\equiv [V^*V^\dag]^{lm}$ is the lepton analogue of the heavy-neutrino tensor $\bm{G}$.

Using a flavour-covariant canonical quantization~\cite{Dev:2014laa}, we may define the matrix number densities of the leptons and heavy neutrinos, as follows:
\begin{subequations}
\begin{align}
  \hspace{-0.85em}\n{L}{s_1 s_2}{p}{l}{m}{t} & \equiv \mathcal V_3^{-1}
  \langle b^m(\ve p, s_2,\tilde{t})
  b_l(\ve p, s_1,\tilde{t})\rangle_t \,,
  \label{eq:def_n_1}\\
  \hspace{-0.85em}\nb{L}{s_1 s_2}{p}{l}{m}{t} & \equiv \mathcal V_3^{-1}
  \langle d_l^{\dagger}(\ve p,s_1,\tilde{t})
  d^{\dagger,m}(\ve p,s_2,\tilde{t})\rangle_t \,,
  \label{eq:def_n_2}\\
  \hspace{-0.85em}\n{N}{r_1 r_2}{k}{\alpha}{\beta}{t} & \equiv
  \mathcal V_3^{-1}
  \langle a^{\beta}(\ve k,r_2,\tilde{t})
  a_\alpha(\ve k,r_1,\tilde{t})\rangle_t \,,
  \label{eq:def_n_3} 
\end{align}
\end{subequations}
where  $\mathcal V_3  = \DiT(\ve  0)$ is the coordinate three-volume and %we have introduced  
the   macroscopic    time
$t=\tilde{t}-\tilde{t}_i$, equal to the   interval  of  microscopic  time
between specification of initial  conditions ($\tilde{t}_i$)
and subsequent observation  of  the  system ($\tilde{t}$)~\cite{Millington:2012pf}. Note the relative reversed ordering of indices in the lepton and anti-lepton number densities, which ensures that the two quantities transform in the same representation, so that they can be combined to form a flavour-covariant lepton asymmetry.  For the Majorana neutrinos, $\bm{n}^N$ and $\overline{\bm{n}}^N$ are not independent quantities and are related by the generalized Majorana condition
\begin{equation}
  \label{eq:Majorana_bar_G}
  \nb{N}{r_1 r_2}{k}{\alpha}{\beta}{t} \
  = \ G_{\alpha \mu} \, \n{N}{r_2 r_1}{k}{\lambda}{\mu}{t} \,
  G^{\lambda \beta} \;.
\end{equation}
The number density matrices defined above have simple generalized-$C$ transformation properties:  
\begin{equation}
[\bm{n}^X(\ve p,t)]^{\widetilde{C}} \ = \ [\overline{\bm{n}}^X(\ve p,t)]^{\mathsf{T}}, 
\label{LN}
\end{equation}
where  $\mathsf{T}$ denotes the matrix transpose acting  on both
flavour and  helicity indices. 
The total number densities $\bm{n}^X(t)$ are obtained by tracing over helicity and isospin indices and integrating over the three-momenta.

Using the $\widetilde{C}$-transformation  relations  \eqref{LN}, we can define the generalized $\widetilde{C}P$-``odd'' lepton asymmetry
\begin{equation}
\bm{\delta n}^L\ = \ \bm{n}^L\:-\:\overline{\bm{n}}^L\;.
\end{equation}
In addition, for the heavy neutrinos, we may define the $\widetilde{C}P$-``even'' and -``odd'' quantities
\begin{equation}
\underline{\bm{n}}^N\ = \ \frac{1}{2}\Big(\bm{n}^N\:+\:\overline{\bm{n}}^N\Big)\;,\qquad \bm{\delta n}^N\ =\ \bm{n}^N\:-\:\overline{\bm{n}}^N\;.
\end{equation}
We will use these quantities, having definite $\widetilde{C}P$-transformation properties, to write down the flavour-covariant rate equations. 

First we derive a Markovian master equation governing the time evolution of the matrix number densities $\mat{n}^X(\ve p,t)$. These are defined in terms of the quantum-mechanical number-density operator 
$\mat{\check{n}}^X(\ve k,\tilde{t};\tilde{t}_i)$  and density operator $\rho(\tilde{t};\tilde{t}_i)$, as follows: 
\begin{equation}
  \label{eq:def_n_rho}
  \mat{n}^{X}(\ve k, t) \equiv \langle
  \mat{\check{n}}^{X}(\ve k,\tilde{t};\tilde{t}_i) \rangle_t 
  = \Tr\left\{\, \rho(\tilde{t};\tilde{t}_i) \,
    \mat{\check{n}}^{X}(\ve k, \tilde{t};\tilde{t}_i) \right\}\,,
\end{equation}
where the trace is over the Fock space. Differentiating \eqref{eq:def_n_rho} with respect to the macroscopic time $t=\tilde{t}-\tilde{t}_i$, and using the Liouville-von Neumann and Heisenberg equations of motion, we proceed via a Wigner-Weisskopf approximation to obtain the leading order Markovian master equation~\cite{Dev:2014laa}
\begin{align}
  \label{eq:master}
  &\frac{\D{}{}}{\D{}{t}} \mat{n}^X(\ve k, t) \
  \simeq \ i  \langle \, [H_0^X,\  \mat{\check{n}}^{X}(\ve k, t) ] \,
  \rangle_t  \nonumber\\& \ -\:\frac{1}{2} \int_{-\infty}^{+\infty} \D{}{t'} \;
  \langle \, [H_{\rm int}(t'),\
  [H_{\rm int}(t),\ \mat{\check{n}}^{X}(\ve k, t)]] \, \rangle_{t} \; ,
\end{align}
where $H^X_0$ and $H_{\rm int}$ are the free and interaction Hamiltonians, respectively.  The first term on the RHS of \eqref{eq:master}, involving the free Hamiltonian, generates flavour oscillations in vacuum, whereas the second term in \eqref{eq:master}, involving the interaction Hamiltonian, generates the collision terms in the generalized Boltzmann equations. 

For the system of lepton and Higgs doublets and heavy-neutrino singlets under consideration, we have  
\begin{subequations}
\begin{align}
  H_0^L & \ = \ \sum_{s}\int_{\ve p} \,
  \Edu{L}{p}{m}{l}
  \Big(b^{m}(\ve p,s,\tilde{t}) \,
  b_l(\ve p,s,\tilde{t}) \; \nonumber \\
   & \qquad \qquad  +\; d^{\dagger}_l(\ve p,s,\tilde{t}) \,
    d^{\dagger,m}(\ve p,s,\tilde{t}) \Big)\; ,
  \label{free_HamL}\\
  H_0^N & \ = \ \sum_{r} \int_{\ve k} \,
  \Edu{N}{k}{\beta}{\alpha} \,
  a^{\dagger,\beta}(\ve k,r,\tilde{t}) \,
  a_\alpha(\ve k,r,\tilde{t}) \; , \label{free_HamN} \\
 H_{\rm int} & \ = \ \int d^4x \h{l}{\alpha} \,
  \bar L^{l} \, \widetilde{\Phi} \, N_{\rm R, \alpha} \,
  + \, {\rm H.c.} \;. \label{Hint}
\end{align} 
\end{subequations}
Using these expressions in \eqref{eq:master}, we obtain the following evolution equations 
for the lepton and heavy-neutrino number densities~\cite{Dev:2014laa}: 
\begin{subequations}
\begin{align}
  &\frac{\D{}{} }{\D{}{t}} \, \n{L}{s_1 s_2}{p}{l}{m}{t} \
  = \ - \, i \,
  \Big[{E}_L(\ve p), \,{n}^{L}_{s_1 s_2}(\ve p,t)
  \Big]_{l}^{\phantom{l}m} \nonumber\\
  &\qquad \qquad  \qquad \qquad 
+ \; [{C}^L_{s_1 s_2}(\ve p,t)]_l^{\phantom l m} \;,
  \label{eq:evol_lept}\\
  &\frac{\D{}{} }{\D{}{t}} \,\n{N}{r_1 r_2}{k}{\alpha}{\beta}{t} \ 
  = \ - \, i \, \Big[{E}_N(\ve k), \,
  {n}^{N}_{r_1 r_2}(\ve k,t)\Big]_{\alpha}^{\phantom{\alpha}\beta} \nonumber\\&\quad
  + \; [{C}^{N}_{r_1 r_2}(\ve k,t)]_\alpha^{\phantom \alpha \beta} 
+ \; G_{\alpha \lambda} \,
  [\overline{{C}}^N_{r_2 r_1}(\ve k,t)]_{\mu}^{\phantom{\mu} \lambda} \,
  G^{\mu \beta}
 \;,
  \label{eq:evol_neu}
\end{align}
\end{subequations}
where, for instance, the lepton collision terms may be written in the form
\begin{align}
  \label{def_coll}
  [{C}^L_{s_1 s_2}(\ve p,t)]_l^{\phantom l m} \ &
  = \ - \: \frac{1}{2} \, \big[ \,{\mathcal{F}} \cdot {\Gamma} \,
  +\, {\Gamma^\dagger} \cdot
  {\mathcal{F}} \,\big]_{s_1 s_2, \,l}^{\phantom{s_1 s_2, \,l} m} \;.
\end{align}
Here,  we   have  suppressed  the  overall   momentum  dependence  and
used a compact notation
\begin{align}
  \label{compact1}
  \big[{\mathcal{F}}
  \cdot {\Gamma} \,\big]_{s_1 s_2, \,l}^{\phantom{s_1 s_2, \,l} m} \ &
  \equiv \ \sum_{s,r_1,r_2} \int_{\ve k, \, \ve q}
  \Tdu{[\mathcal{F}_{s_1 s \,r_1 r_2}
    (\ve p, \ve q, \ve k,t)]}{l}{n}{\alpha}{\beta}
  \nonumber\\&\qquad \times\:
  \Tdu{[\Gamma_{s\, s_2 r_2 r_1}
    (\ve p, \ve q, \ve k)]}{n}{m}{\beta}{\alpha}
  \;.
\end{align} 
In \eqref{compact1}, there are two {\it new rank-4 tensors} in flavour space, as required by flavour-covariance: 
(i) the statistical number density tensors
\begin{align}
  & \Tdu{\mat{\mathcal{F}}(\ve p, \ve q, \ve k,t)}{}{}{}{} \
  = \  n^{\Phi}(\ve q, t) \, \mat{{n}}^L(\ve p, t)  \otimes
  \left[{\mat 1} - \mat{{n}}^{N}(\ve k, t)\right] \nonumber \\
 \ & \qquad  
  - \: \left[1 + n^{\Phi}(\ve q, t)\right] \left[{\mat 1}
    - \mat{{n}}^L(\ve p, t)\right] \otimes  \mat{{n}}^{N}(\ve k, t)\;,
  \label{Fstat}
\end{align}
and (ii) the absorptive rate tensors
\begin{align}
  \Tdu{[\Gamma_{s_1 s_2 r_1 r_2}
    (\ve p, \ve q, \ve k)]}{l}{m}{\alpha}{\beta} & 
  =  \hs{k}{\nu}\;  \h{i}{\lambda}  
  \DiFud{k-p-q}{j}{p}{\mu}{\delta} \notag\\
  &\hspace{-8em}\times \
  \frac{1}{2 E_\Phi(\ve q)}\, \Esud{L}{p}{i}{j} \, \Esdu{L}{p}{k}{n} \,\nonumber \\
& \hspace{-8em}\times \ 
  \Esdu{N}{k}{\lambda}{\mu} \, \Esud{N}{k}{\nu}{\gamma}   \Tr\Big\{\su{r_2}{k}{\delta}{\beta} \, \notag\\
& \hspace{-8em}\times \ 
  \sub{r_1}{k}{\gamma}{\alpha} \, {\rm {P_L}} \,
  \su{s_2}{p}{n}{m} \, \sub{s_1}{p}{p}{l} \, {\rm {P_R}} \Big\}
  \;. \label{eq:Gamma1}
\end{align}

\begin{figure*}[t]
  \centering
\subfloat[][Heavy-neutrino self-energy,\\ $N \to L\Phi \to N$.\label{cuta}]{\includegraphics[scale=0.55]{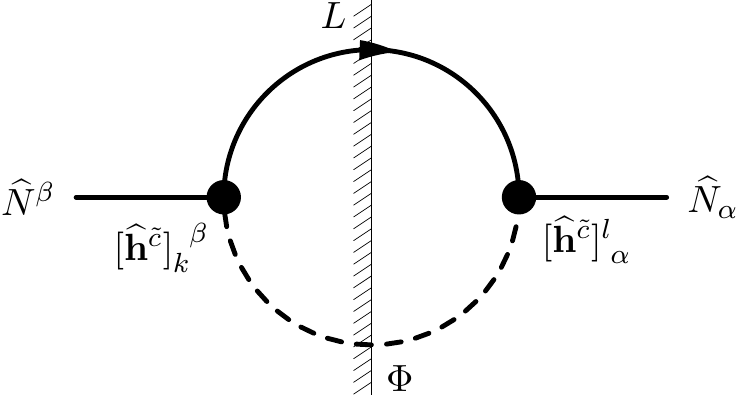}}
\hspace{1.5em}
\subfloat[][Heavy-neutrino production, $n^{\Phi}\protect{[}n^L\protect{]}_l^{\protect\phantom{l}k}\protect{[}\gamma(L\Phi  \to  N)\protect{]}_{k \protect\phantom{l} \alpha}^{\protect\phantom{k} l \protect\phantom{\alpha}\beta}$\label{cutb}]{\includegraphics[scale=0.65]{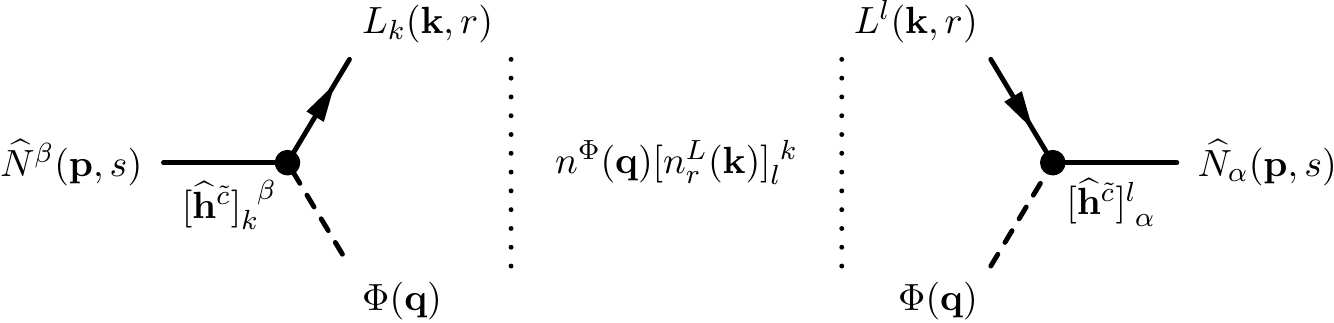}}\\
\subfloat[][Charged-lepton self-energy, with 
        $\Delta L  =  0$ internally.\label{cutc}]{\includegraphics[scale=0.55]{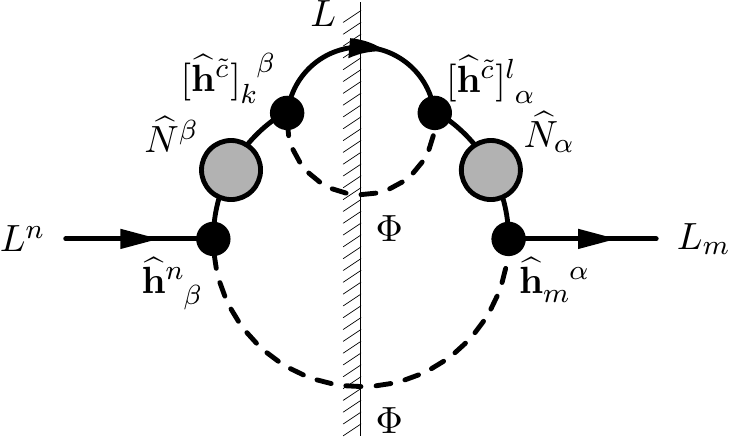}}
\hspace{1.5em}
\subfloat[][$\Delta L = 0$ scattering, $n^{\Phi}\protect{[}n^L\protect{]}_l^{\protect\phantom{l}k}
        \protect{[}\gamma(L\Phi  \to  L\Phi)
        \protect{]}_{k \protect\phantom{l} m}
        ^{\protect\phantom{k} l \protect\phantom{m}n}$.\label{cutd}]{\includegraphics[scale=0.65]{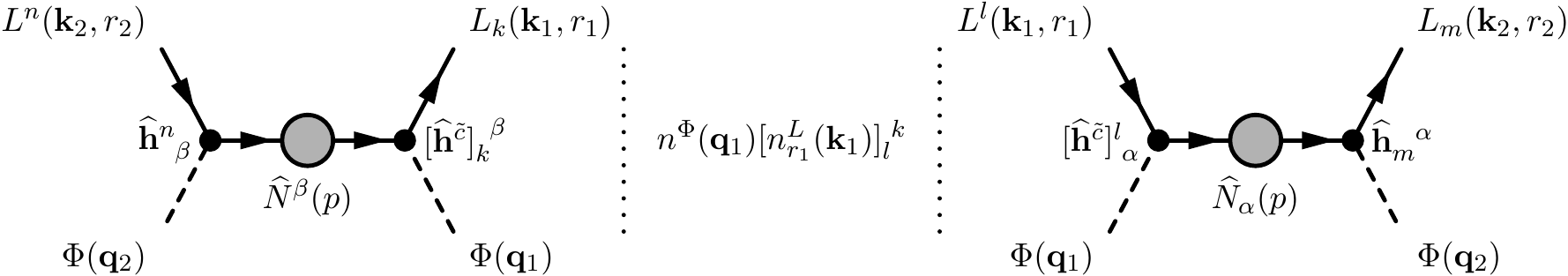}}
\caption{Generalized unitarity cut of the partial heavy-neutrino and lepton self-energies, giving rise to the rank-4 tensor rates for heavy-neutrino production and $\Delta L = 0$ scattering processes. 
The explicit forms of the thermally-averaged rank-4 rates
can be found in~\cite{Dev:2014laa}.
} \label{fig:cuts2}
\end{figure*}

The rate tensor \eqref{eq:Gamma1} describes heavy neutrino decays and inverse decays, and its off-diagonal components are responsible for the evolution of flavour-coherences in the system. The necessary emergence of these higher-rank tensors in flavour space may be understood in terms of the unitarity cuts of the partial self-energies~\cite{Dev:2014laa}. This is illustrated diagrammatically in Figure~\ref{fig:cuts2} for the in-medium heavy-neutrino production $L\Phi \to N$ (Figures~\ref{cuta} and \ref{cutb}) and $\Delta L=0$ scattering $L\Phi\to L \Phi$ (Figures~\ref{cutc} and \ref{cutd}) in a spatially-homogeneous statistical background of lepton and Higgs doublets. In Figures~\ref{cuta} and \ref{cutc},  the cut,  across  which positive  energy  flows  from  unshaded to  shaded
regions, is associated with production rates in the thermal plasma, as
described  by  a generalization of the optical theorem~\cite{Dev:2014laa}. 

%%%%%%%%%%%%%%%%%%%%%%%%%%%%%   %%%%%%%%%%%%%%%%%%%%%%%%%%%%%%%%%%%%%%%

\section{Rate Equations for Resonant Leptogenesis} \label{sec:3}

As already mentioned in  Section~\ref{sec:1}, in  the limit  when two  (or more)  heavy Majorana neutrinos become  degenerate, the $\varepsilon$-type  $ \CP$-violation due to the interference between the tree-level and absorptive part of the self-energy graphs in the heavy-neutrino decay can be resonantly enhanced, even up to order one~\cite{Pilaftsis:1997dr}. In this regime, finite-order perturbation  theory breaks  down and  one needs a consistent field-theoretic resummation of the  self-energy corrections.
Neglecting thermal loop effects~\cite{Giudice:2003jh}, we perform such resummation along the lines of~\cite{Pilaftsis:2003gt} and 
replace the  tree-level  neutrino  Yukawa couplings by  their resummed  counterparts in the  transport equations given  in Section~\ref{sec:2}.  Specifically,  for the  processes $N
\to  L  \Phi$ and  $L^{\tilde{c}}  \Phi^{\tilde{c}}  \to  N$, we  have
$\h{l}{\alpha}  \to  \hr{l}{\alpha}$  and,  for $N  \to  L^{\tilde{c}}
\Phi^{\tilde{c}}$  and $L  \Phi \to  N$, we  have  $\hs{l}{\alpha} \to
\hrc{l}{\alpha}$, where $\tilde{c}$ denotes the $\widetilde{C}\!P$-conjugate.  The algebraic form of the resummed  neutrino Yukawa  couplings in  the heavy-neutrino mass eigenbasis can be found in~\cite{Pilaftsis:2003gt} and the corresponding form in a general flavour basis may be obtained by
the   appropriate  flavour  transformation,   i.e.~$\hr{l}{\alpha}  =
V_{l}^{\ m}U^{\alpha}_{\  \beta}\widehat{\mathbf{h}}_{m}^{\ \ \beta}$,
where                  $\widehat{\mathbf{h}}_{m}^{\                  \
\beta}\equiv\widehat{\mathbf{h}}_{m\beta}$ in the mass eigenbasis~\cite{Dev:2014laa}. 

In  order to  obtain the  rate  equations relevant for RL from  the general  transport
equations         \eqref{eq:evol_lept} and
\eqref{eq:evol_neu}, we perform the following standard approximations: 
\begin{itemize}
\item [(i)] assume kinetic equilibrium, since elastic scattering processes rapidly equilibrate the momentum distributions for all the relevant particle species on time-scales much smaller than their statistical evolution.
\item [(ii)] 
neglect the  mass  splittings  between
different heavy-neutrino flavours inside thermal integrals, and use an average  mass $m_N$ and energy $E_N(\ve k) = (|\ve  k|^2 + m_N^2)^{1/2}$, since the average momentum scale $|\ve k|\sim T \gg |m_{N_\alpha}-m_{N_\beta}|$.
\item [(iii)]
take the classical  statistical limit of \eqref{Fstat}. 
\item [(iv)] neglect thermal and chemical-potential effects~\cite{Pilaftsis:2005rv}. 
\end{itemize}

With the above approximations, we integrate   both  sides  of   \eqref{eq:evol_lept}  and \eqref{eq:evol_neu}, and their generalized $\widetilde{C}P$-conjugates, over the phase space and sum over the degenerate isospin and helicity degrees of freedom. The resulting rate equations account for the decay and inverse decay of  the heavy neutrinos in a flavour-covariant way~\cite{Dev:2014laa}. However, in order to guarantee the correct equilibrium behaviour, we must include the washout terms induced by the  $\Delta L=0$ and $\Delta L=2$ scattering processes, with proper real intermediate state (RIS) subtraction~\cite{Kolb:1980, Pilaftsis:2003gt, Dev:2014laa} (see e.g.,  Figure~\ref{cutd}).  As illustrated in~\cite{Dev:2014laa}, it is necessary to account for thermal corrections in the RIS contributions, when considering off-diagonal flavour correlations. 

In addition to the $2\leftrightarrow 2$ scatterings, it is also important to include the effect of the charged-lepton Yukawa couplings, which are responsible for the decoherence of the charged leptons towards their would-be mass eigenbasis, as opposed to the interactions with the heavy neutrinos [cf.~\eqref{eq:L}], which tend to create a coherence between the charged-lepton flavours. Note
that, while calculating the reaction rates for the processes involving the
charged-lepton Yukawa couplings, it  is important to take into account
their thermal  masses, which control  the phase-space  suppression for
the decay  and inverse  decay of the  Higgs boson~\cite{Cline:1993bd}. 

Taking into account all of these contributions, as well as  the expansion of the Universe, we derive the following  {\it manifestly} flavour-covariant rate equations for the normalized $\gCP$-``even" number 
density matrix $\mat{\underline{\eta}}^{N}$ and $\gCP$-``odd" number density 
matrices $\mat{\delta \eta}^N$ and $\mat{\delta \eta}^L$ (where $\eta^X=n^X/n^{\gamma}$, $n^\gamma$ being the photon number density)~\cite{Dev:2014laa}:
\begin{subequations}
\boxalign[0.45\textwidth]{
\begin{align}
 \frac{H_{N} \, n^\gamma}{z}\,
   &\frac{\D{}{[\underline{\eta}^{N}]_{\alpha}^{\phantom{\alpha}\beta}}}{\D{}{z}} \   = \ - \, i \, \frac{n^\gamma}{2} \,
  \Big[\mathcal{E}_N,\, \delta \eta^{N}\Big]_\alpha^{\phantom \alpha \beta}
+ \, \Tdu{\big[\widetilde{\rm Re}
    (\gamma^{N}_{L \Phi})\big]}{}{}{\alpha}{\beta} \,
 \notag\\
  &
  - \, \frac{1}{2 \, \eta^N_{\rm eq}} \,
  \Big\{\underline{\eta}^N, \, \widetilde{\rm Re}(\gamma^{N}_{L \Phi})
  \Big\}_{\alpha}^{\phantom{\alpha}\beta} \;,
  \label{eq:evofinal2}\\[0.5em]
  \frac{H_{N} \, n^\gamma}{z}\,
  &\frac{\D{}{[\delta \eta^N]_\alpha^{\phantom \alpha \beta}}}{\D{}{z}} \ 
  = \ - \, 2 \, i \, n^\gamma \,
  \Big[\mathcal{E}_N,\, \underline{\eta}^{N}\Big]_\alpha^{\phantom \alpha \beta} \notag\\
  & + \, 2\, i\,  \Tdu{\big[\widetilde{\rm Im}
    (\delta \gamma^{N}_{L \Phi})\big]}{}{}{\alpha}{\beta} \, - \, 
  \frac{i}{\eta^N_{\rm eq}} \, \Big\{\underline{\eta}^N, \,
  \widetilde{\rm Im}
  (\delta\gamma^{N}_{L \Phi}) \Big\}_{\alpha}^{\phantom{\alpha}\beta} \notag\\
  & - \, \frac{1}{2 \, \eta^N_{\rm eq}}  \,
  \Big\{\delta \eta^N, \, \widetilde{\rm Re}(\gamma^{N}_{L \Phi})
  \Big\}_{\alpha}^{\phantom{\alpha}\beta}
  \label{eq:evofinal3}\;, \\[0.5em]
 \frac{H_{N} \, n^\gamma}{z}\, 
& \frac{\D{}{[\delta \eta^L]_l^{\phantom l m}}}
  {\D{}{z}} \ 
  = \ - \, \Tdu{[\delta \gamma^{N}_{L \Phi}]}{l}{m}{}{} \,
  +\, \frac{[\underline{\eta}^{N}]_{\beta}^{\phantom{\beta}\alpha}}
  {\eta^N_{\rm eq}} \,
  \Tdu{[\delta \gamma^{N}_{L \Phi}]}{l}{m}{\alpha}{\beta} \notag\\
  & + \, \frac{[\delta \eta^N]_{\beta}^{\phantom\beta \alpha}}{2\,\eta^N_{\rm eq}} \,
  \Tdu{[\gamma^{N}_{L \Phi}]}{l}{m}{\alpha}{\beta} \, - \frac{1}{3} \,
  \Big\{ \delta {\eta}^{L} , \,
  {\gamma}^{L\Phi}_{L^{\tilde{c}} \Phi^{\tilde{c}}} 
  + {\gamma}^{L\Phi}_{L \Phi}\Big\}_{l}^{\phantom l m} 
 \notag\\
  & 
  \, - \frac{2}{3} \, \Tdu{[\delta {\eta}^L]}{k}{n}{}{} \,
  \Tdu{[{\gamma}^{L\Phi}_{L^{\tilde{c}} \Phi^{\tilde{c}}} - {\gamma}^{L\Phi}_{L \Phi}]}{n}{k}{l}{m} 
 \notag\\[3pt]
  & 
- \frac{2}{3} \, 
  \Big\{\delta \eta^L, \, 
  \gamma_{\rm dec } \Big\}_{l}^{\phantom l m} \,
  +\, [\delta \gamma_{\rm dec}^{\rm back}]_{l}^{\phantom l m}\;.
  \label{eq:evofinal1}
\end{align}}
\end{subequations}
\vspace{0.25em}
Here, $z=m_N/T$, $H_N$ is the Hubble parameter at $z=1$ and $  \mat{\mathcal{E}}_N$ is the thermally-averaged effective heavy-neutrino energy matrix.  $\mat\gamma^N_{L\Phi}$ and $\mat{\delta\gamma}^N_{L\Phi}$ are respectively the $\gCP$-``even" and -``odd" thermally-averaged rate tensors  governing the decay and inverse decay of the heavy neutrinos. The rates $\mat\gamma^{L\Phi}_{L \Phi}$ and $\mat\gamma^{L\Phi}_{L^{\tilde{c}} \Phi^{\tilde{c}}}$ in \eqref{eq:evofinal1} describe the washout due to $\Delta L = 0$ and $\Delta L = 2$ resonant scattering, respectively. On the other hand, $\mat\gamma_{\rm dec}$ and $\mat{\delta \gamma}_{\rm dec}^{\rm back}$ govern the charged-lepton decoherence. An accurate determination of the impact of the latter phenomenon, which tends to suppress the generated asymmetry in models of RL, relies upon the systematic inclusion of the rank-4 absorptive rate tensors. We emphasise, therefore, that this higher-rank flavour structure, which is manifest in this fully flavour-covariant formalism, can play an important physical role.

In obtaining \eqref{eq:evofinal2} and \eqref{eq:evofinal3}, we have defined, for a given Hermitian
matrix  $\mat{A}  =   \mat{A}^\dagger$,  the  generalized  real  and
imaginary parts, as follows:
\begin{subequations}
\begin{align}
  \big[\widetilde{\rm Re}(A)\big]_{\alpha}^{\phantom{\alpha}\beta} \ &
  \equiv \ \frac{1}{2} \, \Big( \Tdu{A}{\alpha}{\beta}{}{} \; 
  + \; G_{\alpha \lambda} \,\Tdu{A}{\mu}{\lambda}{}{}\,
  G^{\mu \beta}\Big) \;, \label{4.26}  \\
  \big[\widetilde{\rm Im}(A)\big]_{\alpha}^{\phantom{\alpha}\beta} \ &
  \equiv \ \frac{1}{2 \, i} \, 
  \Big( \Tdu{A}{\alpha}{\beta}{}{} \; 
  - \; G_{\alpha \lambda} \,\Tdu{A}{\mu}{\lambda}{}{}\,
  G^{\mu \beta}\Big) \;  . \label{4.27}
\end{align}
\end{subequations}
In addition, we have used the relations
\begin{subequations}
\begin{align}
  \widetilde{\rm Re}(\underline{\mat{n}}^N) \ 
  &= \ \underline{\mat{n}}^N \;,\\
  i\, \widetilde{\rm Im}(\mat{\delta n}^N) \ 
  &= \ \mat{\delta n}^N \;.
\end{align}
\end{subequations}

The             flavour-covariant            rate            equations
\eqref{eq:evofinal2}--\eqref{eq:evofinal1} provide a complete and unified  description of the
RL phenomenon,  consistently capturing  the following {\it physically distinct} effects in a  single framework, applicable for any temperature regime:\vspace{-0.5em}
\begin{itemize}
\item [(i)]  Lepton asymmetry due to the {\it resonant mixing}  between heavy neutrinos, as described by the resummed Yukawa couplings in $\mat{\delta\gamma}^N_{L\Phi}$, appearing in the first two terms on the RHS of~\eqref{eq:evofinal1}. This provides a flavour-covariant generalization of the mixing effects discussed earlier in~\cite{Pilaftsis:2003gt}.\vspace{-0.5em}

\item [(ii)] Generation of the lepton asymmetry via coherent heavy-neutrino {\it oscillations}. Even starting with an incoherent diagonal heavy-neutrino number density matrix, off-diagonal $\gCP$-``even'' number densities will be generated at ${\cal O}(h^2)$ due to the $\CP$-conserving part of the coherent inverse decay rate $\mat\gamma^N_{L\Phi}$ in the last two terms on the RHS of~\eqref{eq:evofinal2}. Heavy-neutrino oscillations will transfer these coherences to the $\gCP$-``odd'' number densities $[\delta \eta^N]_\alpha^{\phantom \alpha \beta}$ due to the commutator terms in~\eqref{eq:evofinal2} and \eqref{eq:evofinal3}. Finally, a lepton asymmetry is generated at ${\cal O}(h^4)$ by the $\gCP$-``even'' coherent off-diagonal decay rates in the first term on the second line of~\eqref{eq:evofinal1}. Notice that the novel rank-4 rate tensor $\Tdu{[\gamma^{N}_{L \Phi}]}{l}{m}{\alpha}{\beta}$, required by flavour covariance, plays an important role in this mechanism, along with the $\gCP$-``odd'' number density $[\delta \eta^N]_\alpha^{\phantom \alpha \beta}$, which is purely off-diagonal in the heavy-neutrino mass eigenbasis.  We stress here that this phenomenon of 
coherent oscillations is an ${\cal O}(h^4)$ effect on the {\it total} lepton asymmetry, and so  differs from the ${\cal O}(h^6)$ mechanism proposed in~\cite{Akhmedov:1998qx}. The difference is due to the fact that the latter typically takes place at temperatures much higher than the sterile neutrino masses in the model (see e.g.~\cite{Asaka:2005}), where the total lepton number is not violated at leading order. On the other hand, the ${\cal O}(h^4)$ effect identified here is enhanced in the same regime as the resonant $T=0$ $\varepsilon$-type $ \CP$ violation, namely, for
$z       \approx       1$       and       $\Delta       m_N       \sim
\Gamma_{N_\alpha}$~\cite{Dev:2014laa}.\vspace{-0.5em}

\item [(iii)] {\it Decoherence} effects  due to charged-lepton Yukawa couplings, described by the last two terms on the RHS of~\eqref{eq:evofinal1}. Our description of these effects is similar to the one of~\cite{Abada:2006fw}, which has been generalized here to an arbitrary flavour basis. 
\end{itemize}

%%%%%%%%%%%%%%%%%%%%%%%%%%%%% BD %%%%%%%%%%%%%%%%%%%%%%%%%%%%%%%%%%%%%%%

\section{A Numerical Example} \label{sec:4}
To illustrate the importance of the flavour effects captured {\it only} by the flavour-covariant rate equations \eqref{eq:evofinal2}--\eqref{eq:evofinal1}, we consider a scenario of {\it  Resonant  $\ell$-Genesis}
(RL$_\ell$), in which the final lepton asymmetry is dominantly generated and stored in a {\it single} lepton flavour $\ell$~\cite{Pilaftsis:2004xx}. In  this case, the  heavy neutrino masses could  be as low as the  electroweak scale~\cite{Pilaftsis:2005rv}, still with 
sizable couplings to other  charged-lepton flavours $\ell' \neq \ell$, whilst being  consistent with
all current experimental  constraints~\cite{PDG}.
This     enables     the     modelling     of     RL$_\ell$
scenarios~\cite{Deppisch:2010fr} with electroweak-scale heavy Majorana
neutrinos  that  could be  {\it tested}  during the run-II phase of the  LHC~\cite{Dev:2013wba}. 

The basic assumption underlying RL$_\ell$ models is an approximate SO(3)-symmetric heavy-neutrino sector at some high scale $\mu_X$, with mass matrix $\bm{M}_N(\mu_X)=m_N\bm{1}_3+\bm{\Delta M}_N$~\cite{Pilaftsis:2005rv}. For our purposes, we assume that the SO(3)-breaking mass term $\bm{\Delta M}_N$ is of the minimal form $\bm{\Delta M}_N=\mathrm{diag}(\Delta M_1,\Delta M_2/2,-\Delta M_2/2)$. By virtue of the RG running, an additional mass splitting $\bm{\Delta M}_N^{\mathrm{RG}}$ is induced, such that $\bm{M}_N(m_N)=m_N\bm{1}_3+\bm{\Delta M}_N+\bm{\Delta M}_N^{\mathrm{RG}}$ at the scale relevant to RL. In addition and in order to ensure the smallness of the light-neutrino masses, we require the heavy-neutrino Yukawa sector to have an approximate leptonic U(1)$_l$ symmetry.

As an explicit example of RL$_\ell$, we consider  an
RL$_\tau$  model, with the heavy-neutrino Yukawa structure
\begin{eqnarray}
  \mat{h} \ = \ \left(\begin{array}{ccc}
      0 & ae^{-i\pi/4} & ae^{i\pi/4}\\
      0 & be^{-i\pi/4} & be^{i\pi/4}\\
      0 & ce^{-i\pi/4} & ce^{i\pi/4}
    \end{array}\right) \: + \: \mat{\delta h} \; ,
  \label{yuk}
\end{eqnarray}
where $a,b,c$ are arbitrary complex parameters and
\begin{eqnarray}
  \mat{\delta h} \ = \ \left(\begin{array}{ccc}
      \epsilon_e & 0 & 0\\
      \epsilon_\mu & 0 & 0\\
      \epsilon_{\tau}& 0 & 0
    \end{array}\right)
\label{delta_h}
\end{eqnarray}
contains the U(1)$_l$-breaking parameters $\epsilon_{e,\mu,\tau}$. If the U(1)$_l$ symmetry were to be exact, i.e.~if $\bm{\delta h}=\bm{0}$, then the light neutrino masses would remain massless to all orders in perturbation theory~\cite{Pilaftsis:1991ug}. In order to be consistent with the observed neutrino-oscillation data, we require $|a|,|b|\lsim 10^{-2}$ for electroweak scale heavy neutrinos. In addition, in order to protect the $\tau$ asymmetry from washout effects, we require $|c|\lsim 10^{-5}\ll|a|,|b|$ and $|\epsilon_{e,\mu,\tau}|\lsim 10^{-6}$.

\begin{figure*}[t!]
\centering
\newsavebox{\tempbox}
\hspace{-1.2cm}
\sbox{\tempbox}{\includegraphics[scale=0.55]{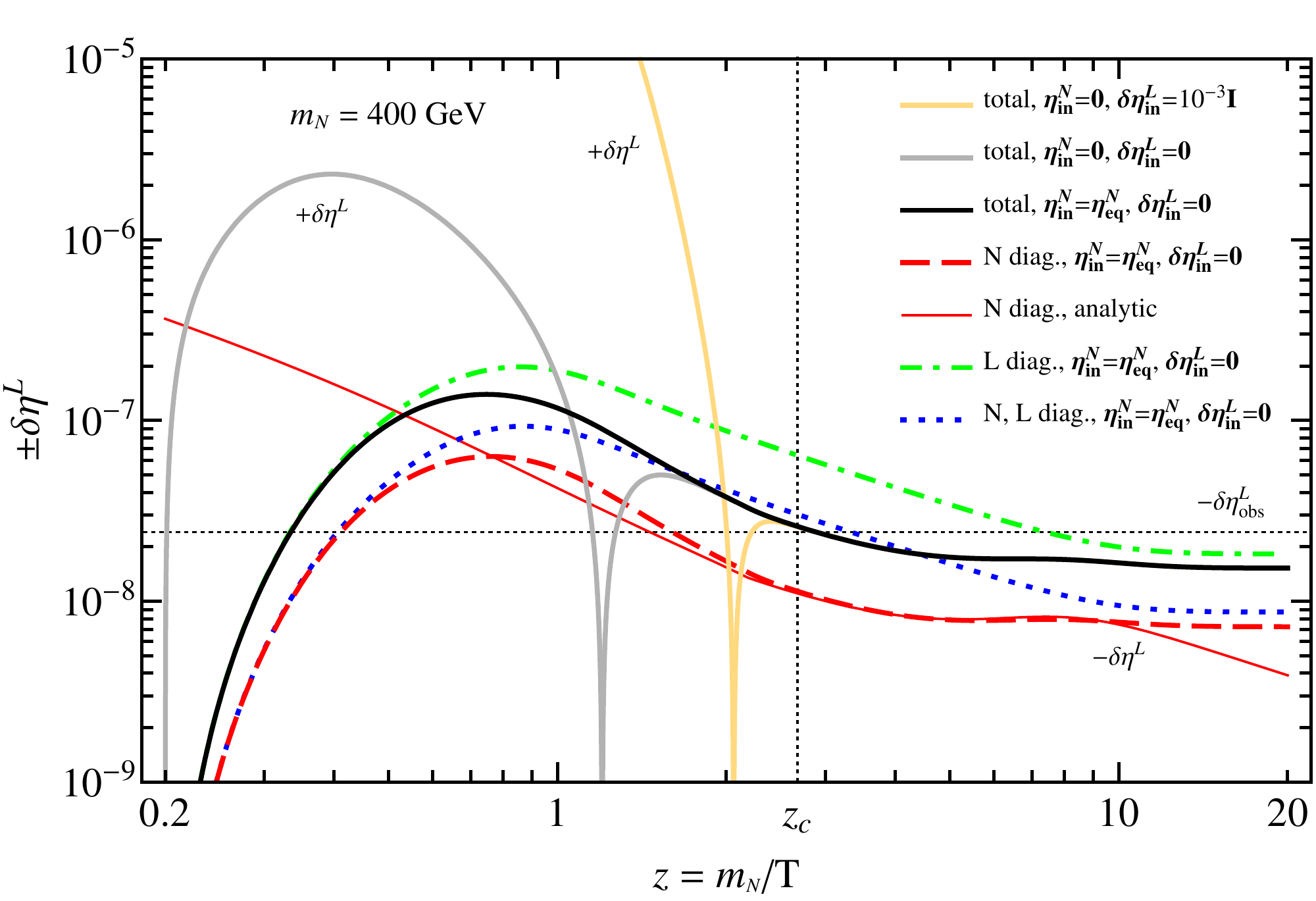}}
\subfloat[ \label{fig2a}]{\includegraphics[scale=0.55]{mN400.pdf}} \hspace{0.5cm}
\subfloat[ \label{fig2b}]{\vbox to \ht\tempbox {\hsize=11em \vfil
\small {\begin{tabular}[b]{c|c}\hline\hline
Parameter & Value \\ \hline\hline
      $m_N$ & 400 GeV \\
      $c$ & $2\times 10^{-7}$ \\
      $\frac{\Delta M_1}{m_N}$ & $-3\times 10^{-5}$ \\ 
      $\frac{\Delta M_2}{m_N}$ & $(-1.21+0.10\,i)\times 10^{-9}$ \\ \hline
      $a$ & 
      $(4.93-2.32 \, i)\times 10^{-3}$ \\
      $b$ &
      $(8.04 - 3.79 \, i)\times 10^{-3}$ \\
      $\epsilon_e$ &
      $5.73\,i\times 10^{-8}$ \\
      $\epsilon_\mu$ &
      $4.30\,i\times 10^{-7}$ \\
      $\epsilon_\tau$ &
      $6.39\,i\times 10^{-7}$ \\  
      \hline\hline
    \end{tabular}}\vfil}}
\caption{(a) Total lepton asymmetry as  predicted by
the RL$_\tau$  model with benchmark parameters given in (b). We show  the
comparison  between  the  total  asymmetry obtained  using  the  fully
flavour-covariant formalism (thick solid lines, with different initial
conditions) with  those obtained using  the flavour-diagonal formalism
(dashed lines).  Also shown (thin  solid line) is an approximate semi-analytic result discussed in~\cite{Dev:2014laa}.}\label{fig2}
\end{figure*}
A choice of benchmark values for these parameters, satisfying all the current experimental constraints, is given in Figure~\ref{fig2b}. The corresponding numerical solution for the total lepton asymmetry  $\delta  \eta^L  \equiv  {\rm  Tr}(\mat{\delta  \eta}^L)$ in our flavour-covariant formalism  is shown in Figure~\ref{fig2a}. Here, the horizontal dotted line 
shows the value of $\delta\eta^L$ required
to  explain  the  observed  baryon  asymmetry in  our  Universe, whereas the vertical line shows the critical temperature $z_c=m_N/T_c$, beyond which the electroweak sphaleron processes become ineffective in converting lepton asymmetry to baryon asymmetry. The thick solid lines show the evolution of $\delta\eta^L$ for three  different initial  conditions, to which the final lepton asymmetry $\delta
\eta^L (z\gg 1)$ is shown to be insensitive. This is a
general consequence of the RL mechanism in the strong washout regime~\cite{Pilaftsis:2005rv}. 

For comparison, we also show in Figure~\ref{fig2a} various partially flavour-dependent limits, i.e.~when either  the heavy-neutrino (dashed  line)  or the lepton (dash-dotted line) number  density   or both (dotted  line) are diagonal in  flavour space. Also shown is the approximate semi-analytic solution discussed in~\cite{Dev:2014laa} for 
the case of a diagonal heavy-neutrino number density (thin solid line).  
The  enhanced lepton asymmetry  in the {\it fully} flavour-covariant  formalism, as compared to when the heavy-neutrino number density is assumed to be diagonal (dashed line), is mainly 
due to the coherent  oscillations  between  the heavy-neutrino
flavours,  leading   to   an
enhancement of a factor of  2.

%%%%%%%%%%%%%%%%%%%%%%%%%%%%%% ALL %%%%%%%%%%%%%%%%%%%%%%%%%%%%%%%%%%%%%

\section{Conclusions}
\label{sec:5}

We have  presented a  {\it fully} flavour-covariant formalism  for transport
phenomena by deriving Markovian master equations that describe the
time-evolution  of  particle   number  densities in  a
quantum-statistical ensemble with  arbitrary flavour content.  As an application, we 
have studied the flavour effects in RL and have obtained {\em manifestly} flavour-covariant rate 
equations for heavy-neutrino  and
 lepton number densities. This provides a complete and unified
description of  RL, capturing  three {\em distinct} physical  phenomena: (i)
resonant  mixing between the heavy-neutrino  states, (ii) coherent
oscillations  between  different heavy-neutrino  flavours and  (iii)
quantum  decoherence   effects  in  the   charged-lepton  sector. The quantitative importance of this formalism and the need to capture consistently all physically-relevent flavour effects has been illustrated for an RL$_\tau$ model. Therein, the
predicted lepton asymmetry is observed to vary by as much as an order of magnitude between the two
partially flavour off-diagonal treatments (dashed and dash-dotted lines in Figure~\ref{fig2a}).

\vspace{-0.5em}

\section*{Acknowledgments}

The work  of P.S.B.D., P.M.  (in part) and  A.P.  is supported  by the
Lancaster-Manchester-Sheffield  Consortium   for  Fundamental  Physics
under  STFC   grant  ST/J000418/1.  P.M.  is also supported in part by the 
IPPP through STFC grant ST/G000905/1.  The work of D.T. has been supported by a fellowship of the EPS  Faculty of  the  University  of Manchester.  

%% The Appendices part is started with the command \appendix;
%% appendix sections are then done as normal sections
%% \appendix

%% \section{}
%% \label{}

%% References
%%
%% Following citation commands can be used in the body text:
%% Usage of \cite is as follows:
%%   \cite{key}         ==>>  [#]
%%   \cite[chap. 2]{key} ==>> [#, chap. 2]
%%

%% References with BibTeX database:
\nocite{*}
\bibliographystyle{elsarticle-num}

%% Authors are advised to use a BibTeX database file for their reference list.
%% The provided style file elsarticle-num.bst formats references in the required Procedia style

%% For references without a BibTeX database:

\end{document}